\documentclass[nonacm,authorversion,screen]{acmart}
\AtBeginDocument{%
  }

\setcopyright{none}

\acmISBN{978-1-4503-XXXX-X/18/06}

\usepackage[normalem]{ulem}
\useunder{\uline}{\ul}{}
\usepackage{multirow}
\usepackage{graphicx}
\usepackage{colortbl}
\usepackage{color}
\usepackage{tabularray}
\definecolor{GuardsmanRed}{rgb}{0.796,0,0}
\definecolor{Gallery}{rgb}{0.937,0.937,0.937}

\begin{document}

\title{Towards an Operational Responsible AI Framework for Learning Analytics in Higher Education}

\author{Alba Morales Tirado}
\email{alba.morales-tirado@open.ac.uk}
\orcid{0000-0001-6984-5122}
\author{Paul Mulholland}
\email{paul.mulholland@open.ac.uk}
\orcid{0000-0001-6598-0757}
\author{Miriam Fernandez}
\email{miriam.fernandez@open.ac.uk}
\orcid{0000-0001-5939-4321}
\affiliation{%
  \institution{Knowledge Media Institute, The Open University}
  \city{Milton Keynes}
  \country{United Kingdom}
}


\begin{abstract}

    Universities are increasingly adopting data-driven strategies to enhance student success, with AI applications like Learning Analytics (LA) and Predictive Learning Analytics (PLA) playing a key role in identifying at-risk students, personalising learning, supporting teachers, and guiding educational decision-making. However, concerns are rising about potential harms these systems may pose, such as algorithmic biases leading to unequal support for minority students. While many have explored the need for Responsible AI in LA, existing works often lack practical guidance for how institutions can operationalise these principles. In this paper, we propose a novel Responsible AI framework tailored specifically to LA in Higher Education (HE). We started by mapping 11 established Responsible AI frameworks, including those by leading tech companies, to the context of LA in HE. This led to the identification of seven key principles such as transparency, fairness, and accountability. We then conducted a systematic review of the literature to understand how these principles have been applied in practice. Drawing from these findings, we present a novel framework that offers practical guidance to HE institutions and is designed to evolve with community input, ensuring its relevance as LA systems continue to develop.

\end{abstract}

\ccsdesc[500]{Applied computing~Learning management systems}

\keywords{Responsible Artificial Intelligence, Learning Analytics, Higher Education}


\maketitle

\section{Introduction}\label{sec:introduction}
Learning Analytics (LA)\footnote{When we refer to Learning Analytics in this paper we consider also Predictive Learning Analytics and any other Learning Analytics approaches supported by Artificial Intelligence} are becoming increasingly central to higher education institutions worldwide. LA systems utilise data to identify at-risk students, support student development, provide personalised and timely feedback, support self-reflection, and enhance the quality of learning and teaching \cite{romero2020educational, sghir2023recent}.\footnote{LA definition \url{https://www.solaresearch.org/about/what-is-learning-analytics/}}

However, the adoption of AI-powered systems within Higher Education (HE) brings with it a range of ethical concerns. Issues like algorithmic bias, lack of transparency, and potential misuse can have serious implications. For example, systems that automatically identify students at risk may suffer from algorithmic biases and disproportionally under-detect students from certain minority groups, leading to those students not receiving equivalent support to the majority group \cite{bayer2021learning}. Other concerns include the need for these automated systems to be able to explain their decisions or the safe and transparent usage of student data. Addressing these issues is crucial to ensure that AI technologies are deployed in a manner that is fair, equitable, and responsible.

In response to the need of addressing the ethical concerns of AI deployment, tech companies and other organisations have developed Responsible AI frameworks to guide the design and development of AI. These frameworks provide guiding principles such as fairness, transparency, accountability, and data privacy to ensure that AI systems are built in a way that minimises harm and maximises societal benefit. However, while Responsible AI frameworks are essential for guiding the design, development and deployment of AI technologies, these frameworks are frequently designed following high-level concepts and principles that can be applied to any AI application, without considering the specificities of the technology or the environment in which the AI system will be deployed. Similarly, numerous works have emerged from the LA community emphasising the importance of ensuring that LA systems adhere to ethical principles \cite{rets2023six, Cerratto-Pargman_McGrath_2021, ethics-australian-hei_2019, ferguson2019ethical}. However, these principles are rarely discussed in practical terms, leaving Higher Education Institutions without clear guidance on how to operationalise them effectively \cite{barletta2023rapid, Tzimas_Demetriadis_2021}. 

To address this gap, this paper introduces a novel Responsible AI framework specifically designed for Learning Analytics (LA) in HE. To develop this framework, we first analysed eleven established Responsible AI frameworks, including those from leading technology companies like Google and Microsoft, and mapped them to the context of LA in HE. From this analysis, we identified seven common Responsible AI principles including: fairness and bias, transparency, privacy, accountability, explainability, safety and security. We then conducted a systematic literature review to explore how existing studies addressed these principles in practice within the LA domain. By synthesising the solutions, challenges, and lessons learned from these studies, we propose a new framework aimed at guiding HE institutions in the responsible implementation of LA systems. Our study is motivated by the following questions: 

\begin{itemize}
    \item {RQ1:  To what extent do existing Responsible AI frameworks address the specific needs and challenges of Learning Analytics in Higher Education?}
    \item {RQ2: Which Responsible AI principles from existing frameworks are applicable to the context of Learning Analytics in Higher Education?}
    \item {RQ3: How have previous Learning Analytics studies incorporated or addressed Responsible AI principles in practice?}
\end{itemize}

Building on the answers to our research questions, we propose the development of a tailored Responsible AI framework for LA in HE. This framework is designed to offer practical, actionable guidance to HE institutions, addressing the ethical, legal, and social complexities inherent in LA systems. We envision this framework as a dynamic resource that evolves with the community, incorporating real-world examples of LA implementations and continuously adapting based on shared challenges and lessons learned. We believe this framework will serve as a valuable asset to the LA community, providing a much-needed tool for operationalising Responsible AI principles in practical ways.




\section{Motivation}\label{sec:motivation}
Higher Education Institutions (HEIs) face several ethical challenges when integrating Artificial Intelligence (AI) into their operations, teaching, research, and administrative functions. These challenges stem from the complexity of AI technologies, the sensitivity of academic and student data, and the societal implications of widespread AI use in education. The spectrum of technologies used is also broad \textcolor{blue}{\cite{Crompton_Burke_2023}}, from Generative AI applications that help to generate new curricula, to AI that can monitor attendance, to LA solutions that could identify students at-risk.

LA systems in particular introduce numerous ethical challenges, especially given their growing use in HE to enhance student success and institutional efficiency. These challenges often stem from the use of vast amounts of student data and algorithmic predictions that can impact decision-making in educational settings. Below is a brief discussion of some of the key ethical issues associated with the use of LA in HE. For a broader overview of the problem, the reader is directed to the following literature \cite{slade2013learning, ferguson2019ethical, roberts2017ethical, mathrani2021perspectives, Circleu_2022, nguyen2023ethical}.

One key ethical issue in LA systems is bias. For instance, a predictive model might unfairly classify students from certain demographics as at-risk or not, leading to unequal treatment and opportunities. If LA decisions guide resource allocation or interventions, some students may receive more support, while others are overlooked. LA systems often inherit bias from the training data \cite{ntoutsi2020bias}, reinforcing social inequalities. Bias can also be introduced during data processing, resulting in different levels of support for students based on factors like race, gender, disability, or socioeconomic status \cite{bayer2021learning}. It is also important to consider that the predictions generated by LA tools can bias the behaviour of students or instructors (e.g, by deciding not to continue a course if the prediction is negative in the case of students, or to grade a student unfairly by over-relying on the output of an algorithmic system) \cite{tsai2021more}. This shows that biases in LA systems are both technical and social, highlighting the need for Responsible AI principles that address both dimensions.

Even when the LA systems are not biased, they are never 100\% right, they make mistakes. LA systems often fail to consider the full context of a student's life, such as personal struggles, cultural differences, or family obligations, which can affect academic performance but are not easily captured by quantitative data \cite{hlosta2022predictive}. Those inaccurate predictions can lead to students either receiving unnecessary interventions or being deprived of necessary support. It is also necessary to reskill staff to the use of LA systems, ensuring these systems are properly understood and used as intended.

Many LA systems function as "black-boxes" \cite{nguyen2023ethical}, meaning their decision-making processes are not easily interpretable. This may lead to mistrust among students and staff. If students don't understand how LA decisions that affected them are made they may feel mistreated. Similarly, if staff does not understand the reasoning behind LA decisions, they may decide not to use them, which could lead to disadvantaging their students.

It is also unclear where the failure lies when the LA system makes an error, and that error leads to harm.\footnote{\url{https://www.wired.com/story/alevel-exam-algorithm/}} Ensuring accountability in LA systems is difficult as they involve various stakeholders, including data scientists, administrators, faculty, third-party vendors, and students. With so many parties involved, it can be difficult to determine who is ultimately accountable for decisions made based on LA insights. For the same reason, when biases and predictive errors occur it is challenging to hold a specific party accountable.

LA systems are based on the analysis of vast amounts of data on students, and staff, including academic records, attendance, or behavioural data from learning management systems. Students may not be fully aware of which data is being collected, for which purposes, and whether it is being shared, as there may be inadequate consent mechanisms. Students may also not have the option to opt out of being monitored by LA systems, which raises ethical questions about autonomy and the right to control one’s own data. It is also often unclear who owns the data — the institution, the student, or even third-party providers of analytics platforms. This lack of clarity can lead to disputes about how data can be used. Also, HE institutions are often targets of cyber attacks\footnote{\url{https://www.gov.uk/government/statistics/cyber-security-breaches-survey-2023/cyber-security-breaches-survey-2023-education-institutions-annex}}, increasing the risk of sensitive information from students and staff being exposed or misused. 

The use of LA systems by different HE stakeholders can also constitute data misuse. It is often unclear within HE institutions which roles should have access to which type of data, and when. Following the principle of data minimisation, staff members should have access to the minimum amount of data required to do their jobs. However, this requires role adaptations of the different LA systems, which are not always possible, specially if the LA solution is acquired from a third-party vendor. 

Concerns have been raised about the psychological impact of LA on both students and staff \cite{roberts2016student}. LA systems can create a sense of constant surveillance, where individuals feel closely monitored. This awareness can affect how staff support students, as teachers might adjust their methods to align with data-driven insights, potentially limiting their teaching style. Similarly, students may limit their creativity and exploration if LA systems restrict their learning options through algorithmic recommendations. This can undermine students' freedom to explore diverse knowledge paths.

Institutions may see LA as a cost-cutting solution. For example, using LA to streamline the effectiveness of academics may lead to job losses or reduce the number of hours within contracts. How and where institutions invest those savings from the use of AI-driven applications also constitutes an ethical dilemma. It is also important to acknowledge the digital divide. As more LA solutions are put in place benefiting both, students and staff, students without access to technology may be at a disadvantage. Similarly, institutions in wealthier countries or regions may have more resources to implement LA solutions effectively, leading to a gap between global education systems.

 In addition, many regions in the world are still in the process of developing comprehensive legal frameworks to govern AI \cite{alvarez2024policy}. HE institutions therefore need to stay up to date with local, national and international regulations. Also, since different regions may have varying legal standards it makes it difficult for institutions with international students to apply consistent ethical standards. These challenges show the importance of creating a practical framework that could guide HE institutions towards the design, development, deployment and use of their LA solutions.

\section{Analysing Existing Responsible AI Frameworks}\label{sec:existingRAI}
To address our first two research questions: (i) \textit{RQ1: To what extent do existing Responsible AI frameworks address the specific needs and challenges of Learning Analytics in Higher Education?} and (ii) \textit{RQ2: Which Responsible AI principles from existing frameworks are applicable to the context of Learning Analytics in Higher Education?}, we followed a comparative analysis of eleven well-established Responsible AI frameworks.

Our methodology began by clearly defining our research objectives and scope. To ensure a comprehensive and diverse set of frameworks, we conducted an extensive search using multiple queries through Google's search engine. Key search terms included ‘Responsible AI framework,’ ‘ethical AI adoption,’ ‘ethical AI in education,’ and ‘responsible AI in education.’ For each query, we reviewed the top 30 results to capture a broad spectrum of frameworks across industry, government, and the education sector. Additionally, we integrated findings from recent literature reviews of Responsible AI frameworks \cite{barletta2023rapid, Tsai_Gasevic_2017}, ensuring the inclusion of widely recognised frameworks from leading technology companies like Microsoft, Amazon, and Google.

Given that many Responsible AI frameworks are not published as traditional academic papers, but rather proposed by industry, government, or third-sector organisations, we opted not to limit our search to scholarly databases. This approach allowed us to capture the most relevant and practical frameworks beyond academic literature.

For inclusion in our analysis, we applied the following eligibility criteria: 
\begin{itemize} 
    \item Documents must be written in English. 
    \item Frameworks must address the ethical use of data or software development practices, considering the ethical, legal, and social challenges related to AI design, development, and adoption. 
    \item Documents describing policies, guidelines, or codes of practice, rather than full frameworks, were included if: (i) they specifically targeted the education sector, or (ii) they were produced by leading technology companies. 
\end{itemize}

This systematic process ensures a robust and diverse dataset, allowing for a thorough analysis of Responsible AI principles relevant to Learning Analytics in Higher Education. As detailed in Table~\ref{tab:compilation-known-frameworks} we identified eleven relevant initiatives launched by organisations in different domains. The table details: (i) the originating organisation, (ii) the document's name and URL, (iii) the primary focus of the document (Artificial Intelligence, Machine Learning, Data, Learning Analytics, or Predictive Learning Analytics), (iv) the document type (framework, policy, principles, code of practice, or guidance), (v) the year of release, (vi) the number of Responsible AI principles discussed, and (vii) the context (domain, country, sector) in which the principles are applied. Although marginally relevant, we excluded the SHEILA Framework \cite{Tsai_Gasevic_2017}, as it did not fully meet our eligibility criteria, being primarily focused on strategic planning and policy processes for Learning Analytics.

\begin{table}[]
\centering
\caption{Responsible AI initiatives: frameworks, guidelines, policies. AI = Artificial Intelligence, ML = Machine Learning, LA = Learning Analytics, PLA = Predictive Learning Analytics. }
\label{tab:compilation-known-frameworks}
\resizebox{\textwidth}{!}{%
\begin{tabular}{cllcllcl}
\hline
\textbf{\#} & \multicolumn{1}{c}{\textbf{Organisation name}} & \multicolumn{1}{c}{\textbf{Document name}} & \textbf{Focus} & \multicolumn{1}{c}{\textbf{Doc. Type}} & \multicolumn{1}{c}{\textbf{Year}} & \textbf{\begin{tabular}[c]{@{}c@{}}Number of \\ Principles\end{tabular}} & \multicolumn{1}{c}{\textbf{Context}} \\ \hline
1 & NIST & AI Risk Management Framework (NIST)~\cite{Tabassi-NIST_2023} & AI & Framework & 2023 & 7 & \begin{tabular}[c]{@{}l@{}}Non-sector-specific\\ AI risk-oriented (design, \\  development, release, and use)\end{tabular} \\ \hline
2 & Microsoft & Microsoft Responsible AI Standard~\cite{microsoft_2022} & AI & Framework & 2022 & 6 & \begin{tabular}[c]{@{}l@{}}AI in industry\\ Product development\end{tabular} \\ \hline
3 & \begin{tabular}[c]{@{}l@{}}The Institute for Ethical\\ AI in Education - IEAIE\end{tabular} & The Ethical Framework for AI in Education~\cite{ieaie_2021} & AI & Framework & 2021 & 9 & \begin{tabular}[c]{@{}l@{}}UK\\ AI in Education\end{tabular} \\ \hline
4 & Alan Turing Institute & \begin{tabular}[c]{@{}l@{}}Understanding artificial intelligence ethics and \\ safety: A guide for the responsible design and \\ implementation of AI systems in the public sector~\cite{Lesli-alan-turing_2019}\end{tabular} & AI & Guideline & 2019 & 4 & \begin{tabular}[c]{@{}l@{}}UK\\ AI in the public sector\end{tabular} \\ \hline
5 & The Open University & Data Ethics Policy - OU (2023)~\cite{ou_policy_2023} & \begin{tabular}[c]{@{}c@{}}AI, ML, \\ Data\end{tabular} & Policy & 2023 & 4 & \begin{tabular}[c]{@{}l@{}}UK. The Open University\\ Data management\end{tabular} \\ \hline
6 & IBM & AI ethics at IBM~\cite{IBM} & AI & Principles description & 2024 & 5 & \begin{tabular}[c]{@{}l@{}}AI in industry\\ Product development\end{tabular} \\ \hline
7 & Google & Google’s AI Principles~\cite{Google_2023} & AI & Principles description & 2023 & 7 & \begin{tabular}[c]{@{}l@{}}AI in industry\\ Product development\end{tabular} \\ \hline
8 & Amazon & Building AI responsibly at AWS~\cite{amazon_na} & AI & Principles description & NE & 8 & \begin{tabular}[c]{@{}l@{}}AI in industry\\ Product development\end{tabular} \\ \hline
9 & JISC & Code of practice for learning analytics~\cite{jisc_code_2023} & LA & Code of practice & 2023 & 7 & \begin{tabular}[c]{@{}l@{}}UK \\ Learning Analytics in HEI\end{tabular} \\ \hline
10 & \begin{tabular}[c]{@{}l@{}}ICDE - International \\ Council for Open and \\ Distance Education\end{tabular} & Global Guidelines: Ethics in Learning Analytics~\cite{Slade_Tait_icde-2019} & \begin{tabular}[c]{@{}c@{}}LA, \\ PLA\end{tabular} & Guideline & 2019 & 10 & \begin{tabular}[c]{@{}l@{}}Global\\ Learning Analytics \\ in Education\end{tabular} \\ \hline
11 & University of Edinburgh & Learning Analytics Principles and Purposes~\cite{UoE_2017} & LA & Policy & 2017 & 7 & \begin{tabular}[c]{@{}l@{}}UK. University of Edinburgh\\ Learning Analytics in HEI\end{tabular} \\ \hline
\end{tabular}%
}
\end{table}

We then conducted a \textit{SWOT (Strengths, Weaknesses, Opportunities, and Threats)} analysis of the eleven selected documents. The Strengths criterion evaluated how well each document addressed the unique needs and challenges of Learning Analytics in Higher Education. Weaknesses identified areas where the documents could be improved to better meet these needs. Opportunities considered how the insights and lessons from each document could contribute to developing a tailored Responsible AI framework for LA in HE. Finally, Threats examined any potential challenges or competition highlighted by the documents that could impact the development of such a framework.

Our analysis revealed that most frameworks adopt a holistic view of AI systems, often focusing on machine learning algorithms and large-scale data science without specific consideration of LA in HE. While some documents, such as those by \cite{jisc_code_2023, Slade_Tait_icde-2019, UoE_2017, ou_policy_2023}, provide relevant high-level principles and guidelines, \textbf{they often lack concrete, actionable steps, tools, or measurable practices that HE institutions could implement in their LA efforts}.

However, we identified significant opportunities within these frameworks. Many of the principles outlined, if refined and tailored specifically to the LA context in HE, could serve as a solid foundation for the development of a robust and applied Responsible AI framework. We then proceeded to analyse the Responsible AI principles mentioned across these documents, extracting their commonalities and unique aspects. This allowed us to identify key principles that could be adapted to meet the needs of LA in HE and inform the creation of our framework.

Table~\ref{tab:common-principles} presents the common Responsible AI principles identified across the 11 analysed documents. Each row corresponds to one of the analysed documents and highlights the Responsible AI principles discussed within it. It is important to note that while different documents may use slightly varied terminology, they often refer to the same underlying principles. In the top row, we list the common Responsible AI principles identified across all documents, along with the standardised names we have selected for each principle in this study. Subsequent rows display the individual documents, indicating which of the common principles they cover and how those principles are named in the respective document. Note that certain documents may group two or more principles under a common one (e.g., Transparency and accountability \cite{Slade_Tait_icde-2019} or Privacy and Security \cite{microsoft_2022}). In those cases, the principle appears twice in the table under the different individual ones. The final column, "Unclassified," includes principles that are either unique to a specific policy, guideline, or framework, or that appear only in a small subset of the analysed documents.

Building on the various definitions of common principles found in the 11 analysed documents, we propose here a set of definitions specifically tailored to the context of Learning Analytics in the Higher Education sector:
\vspace{-0.145cm}
\begin{itemize}
    \item {\textbf{Fairness and Bias}: Ensuring that LA systems are free from biases that could disadvantage certain student groups, such as minorities or underrepresented communities. Fairness in this context means that predictive models do not disproportionately label or classify students based on sensitive attributes like gender, race, or socio-economic background, that any interventions derived from analytics are equitably distributed among all students, and that the outputs of LA systems and are not biasing students and staff in their decisions.}
    \item {\textbf{Transparency}: Providing clear and accessible information to all stakeholders (students, educators, administrators) about how LA systems operate, including data collection, algorithms used, and the decision-making processes behind predictive analytics. This principle emphasises the importance of openness in the system's design, implementation, and outcomes, ensuring that users understand how predictions and classifications are made and which and how their data is being utilised.}
    \item {\textbf{Accountability}: Establishing clear lines of responsibility for the design, deployment, and outcomes of LA systems. This includes holding institutions, technology providers, and stakeholders accountable for ensuring ethical practices, addressing unintended consequences, and mitigating harms caused by LA-driven decisions. In the context of Higher Education, accountability particularly ensures that institutions take responsibility for the accuracy and fairness of predictive analytics outcomes and their impact on students.}
    \item {\textbf{Privacy}: Protecting the personal data of students and staff, ensuring that LA systems comply with legal and ethical standards around data privacy, such as GDPR.\footnote{\url{https://gdpr-info.eu/}} This includes collecting only necessary data, securely storing it, and ensuring that students and staff have control over how their data is used. Privacy also involves limiting the sharing of personal data to authorised individuals or systems and ensuring that predictive models do not intrude on the personal lives of students or staff.}
    \item {\textbf{Security}: Safeguarding LA systems from data breaches, hacking, and unauthorised access to sensitive student and staff data. This principle focuses on implementing robust technical safeguards to protect the integrity, confidentiality, and availability of data and ensuring that predictive models are secure from manipulation or misuse. Security is particularly important in HE, where large volumes of sensitive information are processed.}
    \item {\textbf{Explainability}: Ensuring that the predictions and decisions made by LA systems can be understood by non-expert users, such as HE staff and students. Explainability in this context involves providing clear, understandable explanations for how specific predictions and decisions were reached and offering insights into the variables that contributed to those outcomes.} 
    \item {\textbf{Safety}: Ensuring that LA systems are designed to minimise harm to staff and students, whether psychological, emotional, or academic. Safety in this context involves evaluating the potential risks of using LA, such as flawed predictions, over-reliance on predictions or biases derived from the human perceptions of those predictions, and ensuring that interventions based on analytics are supportive rather than punitive. It also means ensuring that these systems do not create undue stress or pressure on students and staff.}     
\end{itemize}

Unclassified principles such as Sustainability, Cultural values, or Student agency and Responsibility are not considered in our proposed framework, because they were either less explicitly defined in the context of Learning Analytics or do not directly align with the immediate operational needs and challenges identified in our analysis of existing frameworks. In the following section, our goal has been to review existing works on LA (tools, applications, use cases) and extract valuable lessons learned, including best practices, challenges, opportunities, in relation to the seven identified principles. 



\begin{table}[]
\centering
\caption{Responsible AI common principles identified across analysed documents.}
\label{tab:common-principles}
\resizebox{\textwidth}{!}{%
\begin{tabular}{l|llllllll}
\hline
 & \multicolumn{8}{c}{\textbf{Common principles}} \\ \cline{2-9} 
\multirow{-2}{*}{\textbf{\begin{tabular}[c]{@{}l@{}}Organisation's \\ name\end{tabular}}} & \multicolumn{1}{c}{\textbf{Fairness and bias}} & \multicolumn{1}{c}{\textbf{Transparency}} & \multicolumn{1}{c}{\textbf{Accountability}} & \multicolumn{1}{c}{\textbf{Privacy}} & \multicolumn{1}{c}{\textbf{Security}} & \multicolumn{1}{c}{\textbf{Explainability}} & \multicolumn{1}{c}{\textbf{Safety}} & \multicolumn{1}{c}{\textbf{Unclassified}} \\ \hline
\rowcolor[HTML]{EFEFEF} 
\textbf{\begin{tabular}[c]{@{}l@{}}Alan Turing \\ Institute\end{tabular}} & Fairness & Transparency & Accountability & - & - & - & - & Sustainability \\
\textbf{\begin{tabular}[c]{@{}l@{}}The Open \\ University\end{tabular}} & Fairness & Transparency & Accountability & - & - & Explainability & - & - \\
\rowcolor[HTML]{EFEFEF} 
\textbf{ICDE} & Inclusion & \begin{tabular}[c]{@{}l@{}}- Communications\\ - Transparency\end{tabular} & \begin{tabular}[c]{@{}l@{}}Institutional responsibility  \\ and obligation to act\end{tabular} & \begin{tabular}[c]{@{}l@{}}- Consent\\ - Data ownership \\ and control\end{tabular} & \begin{tabular}[c]{@{}l@{}}- Accessibility of data\\ - Validity and reliability \\ of data\end{tabular} & - & - & \begin{tabular}[c]{@{}l@{}}- Cultural values\\ - Student agency and \\ responsibility\end{tabular} \\
\textbf{Google} & \begin{tabular}[c]{@{}l@{}}- Avoid creating or \\ reinforcing unfair bias\\ - Be socially beneficial\end{tabular} & - & Be accountable to people & \begin{tabular}[c]{@{}l@{}}Incorporate privacy \\ design principles\end{tabular} & - & - & \begin{tabular}[c]{@{}l@{}}- Be built and \\ tested for safety\\- Be made available for \\ uses that accord \\ with these principles\end{tabular} & \begin{tabular}[c]{@{}l@{}}- Uphold high standards of \\ scientific excellence\end{tabular} \\
\rowcolor[HTML]{EFEFEF} 
\textbf{IBM} & Fairness & Transparency & - & Privacy & Robustness & Explainability & - & - \\
\textbf{IEAIE} & Equity & \textit{\begin{tabular}[c]{@{}l@{}}- Informed Participation \\ - Transparency and \\ Accountability\end{tabular}} & \textit{\begin{tabular}[c]{@{}l@{}}- Transparency and \\ Accountability\end{tabular}} & Privacy & - & - & \begin{tabular}[c]{@{}l@{}}- Achieving \\ Educational Goals\end{tabular} & \begin{tabular}[c]{@{}l@{}}- Administration and Workload\\ - Autonomy\\ - Ethical Design\\ - Forms of Assessment\end{tabular} \\
\rowcolor[HTML]{EFEFEF} 
\textbf{JISC} & \begin{tabular}[c]{@{}l@{}}Minimising \\ adverse impacts\end{tabular} & \begin{tabular}[c]{@{}l@{}}- Transparency, legal \\ basis and consent\\ - Access\end{tabular} & Responsibility & \begin{tabular}[c]{@{}l@{}}- Privacy \\ - Stewardship of data\end{tabular} & - Validity & - & - & - \\
\textbf{\begin{tabular}[c]{@{}l@{}}University of \\ Edinburgh\end{tabular}} & \begin{tabular}[c]{@{}l@{}}- Beneficial to students\\ - Fairness and bias\end{tabular} & Transparent & \begin{tabular}[c]{@{}l@{}} - Be accountable \\ to people \\ - Governance \end{tabular} & Privacy & - & Explainability & - & -  \\
\rowcolor[HTML]{EFEFEF} 
\textbf{Microsoft} & \begin{tabular}[c]{@{}l@{}}- Fairness\\ - Inclusiveness\end{tabular} & Transparency & Accountability & \textit{Privacy and Security} & \textit{\begin{tabular}[c]{@{}l@{}}- Privacy and \\ Security\end{tabular}} & - & \begin{tabular}[c]{@{}l@{}}Reliability \\ and Safety\end{tabular} & - \\
\textbf{Amazon} & Fairness & Transparency & - Governance & \textit{Privacy and security} & \textit{\begin{tabular}[c]{@{}l@{}}- Privacy and security\\ - Veracity and robustness\end{tabular}} & \begin{tabular}[c]{@{}l@{}}- Explainability\\ - Controllability\end{tabular} & Safety & - \\
\rowcolor[HTML]{EFEFEF} 
\textbf{NIST} & \begin{tabular}[c]{@{}l@{}}Fair with harmful \\ bias managed\end{tabular} & \textit{\begin{tabular}[c]{@{}l@{}}- Accountable and \\ transparent\end{tabular}} & \textit{\begin{tabular}[c]{@{}l@{}}Accountable and \\ transparent\end{tabular}} & Privacy-enhanced & \begin{tabular}[c]{@{}l@{}}- Secure and resilient\\ - Valid and reliable\end{tabular} & \begin{tabular}[c]{@{}l@{}}Explainable \\ and interpretable\end{tabular} & - Safe & - \\ \hline
\end{tabular}%
}
\end{table}

\section{Analysing LA works with respect to Responsible AI principles}\label{sec:responsibleAIPrinciples}
We address in this section the third research question \textit{(RQ3): How have previous Learning Analytics studies incorporated Responsible AI principles in practice?} To answer this, we conducted a systematic literature review of relevant studies.

\subsection{A Systematic Literature Review}

We initiated our systematic literature review by identifying key terms derived from our research questions and the ethical principles discussed (see Table~\ref{tab:common-principles}). We compiled a list of synonyms to create a comprehensive search string using Boolean operators (AND, OR). The structured search query was formulated as follows: \{domain of interest\} + \{area of implementation\} + \{principles\} + \{focus\}. The resulting search string was defined as \textit{('learning analytics' OR 'predictive learning analytics') AND ('higher education') AND (fairness OR transparency OR privacy OR accountability OR safety OR explainability OR ethics OR 'responsible AI') AND (framework OR guideline OR policy OR 'code of practice' OR principles OR 'best practice' OR implications OR 'lessons learn')}. 

This search string was applied across three digital libraries—ERIC \url{https://eric.ed.gov/}, SCOPUS \url{https://www.scopus.com}, and ACM \url{https://dl.acm.org} —selected for their relevance to our study. Searches were conducted on titles, abstracts, and keywords. We obtained: ERIC (54), Scopus (70) and ACM (110) results from each library. Before selection, we removed duplicate results. Subsequently, we established robust inclusion and exclusion criteria based on our research questions (see Table~\ref{tab:literature-criteria}). We focused on studies that explore the responsible adoption of LA. This included papers detailing lessons learned from LA implementations, identifying challenges faced by Higher Education Institutions (HEIs), and proposing strategies to address these issues, including policy and guideline development.


\begin{table}[]
\small
\centering
\caption{Inclusion and exclusion criteria}
\label{tab:literature-criteria}
\begin{tabular}{ll}
\hline
\multicolumn{2}{l}{\textbf{Inclusion criteria}}                \\ \hline
1 & Studies that describe ethical concerns in the adoption LA or PLA by HEI \\
2 & Studies that describe ethical challenges or address specific ethical principles for adoption of LA  \\
3 & Studies that discuss, suggest or have implemented controls, guidelines or policies for ethical adoption of LA.  \\ \hline
\multicolumn{2}{l}{\textbf{Exclusion criteria}}                \\ \hline
1 & Studies written in a language other than English           \\
2 & Conference abstracts and editorials                        \\
3 & Studies that do not meet any of the inclusion criteria     \\ 
4 & Studies that focus on other education organisations other than HEI \\ \hline
\end{tabular}
\end{table}

The study selection followed a three-stage process: (a) reviewing the titles and abstracts, (b) reading the introductions and conclusions, and (c) evaluating the full text. At each stage, documents were categorised into three groups: 'important,' 'unsure,' and 'not relevant.' Papers classified as 'unsure' were reviewed collectively by all authors to reach a final decision. Additionally, we employed a snowballing technique to manually include certain papers identified through key citations.

A total of 234 articles were initially identified through database searches. After removing duplicates (N=37), 197 articles proceeded to the three-stage selection process. Ultimately, 167 studies were excluded, and 30 were selected. An additional 15 studies were manually added, resulting in a final total of 45 studies for this literature review. 

\subsection{Findings}

The publication years of the selected papers range from 2013 to 2024. 
The selected studies include journal articles, book chapters, and conference papers.
We first classified the selected works according to the seven principles identified in Section~\ref{sec:existingRAI}. The classification was done in two steps: (a) reading the titles and abstracts to identify the potential Responsible AI principle(s) addressed by each study, and (b) reviewing the full text, with a focus on the methodology, results, and discussion sections, to determine the primary and secondary principles covered. As shown in Table~\ref{tab:papers-by-principle}, 40\% of the analysed studies primarily focus on Privacy. The second most commonly addressed principle is Transparency. We also created a 'Various principles' category for works focusing on more than two principles. In the following section, we review the selected works, discussing how they have applied Responsible AI principles in practice, the challenges encountered, and the lessons learned.

\begin{table}[h!]
\small
\caption{Distribution of Studies Across Responsible AI Principles}
\label{tab:papers-by-principle}
\centering
\begin{tabular}{lcccccccc}
\hline
\textbf{Focus}       & \textbf{Privacy} & \textbf{Transparency} & \textbf{Fairness/Bias} & \textbf{Accountability} & \textbf{Safety} & \textbf{Security} & \textbf{Explainability} & \textbf{Various} \\ \hline
Primary              & 18               & 9                     & 7                      & 3                       & 1               & 4                & 1               & 2        \\
Secondary            & 5                & 6                     & 2                      & 0                       & 2               & 0                & 0             & 6          \\ \hline
\end{tabular}
\end{table}

\subsubsection{Accountability}

The accountability principle mandates that institutions take responsibility for decisions generated by predictive analytics systems. All stakeholders—such as HEI directors, managers, and data scientists—must understand their roles throughout the lifecycle of Learning Analytics (LA) systems. For instance,  \cite{Patterson_York_Maxham_Molina_Mabrey_2023} highlight the importance of accountability in the design phase, particularly concerning data management. Similarly, \cite{Alzahrani_Tsai_Aljohani_Whitelock-wainwright_Gasevic_2023} emphasise that unclear governance policies can undermine trust in LA systems.
\cite{Veljanova_Barreiros_2022} provide operational criteria for accountability, such as creating clear documentation of roles for developers and users of LA dashboards. Compliance with GDPR also plays a critical role, in defining key responsibilities among data controllers, processors, and subjects.
The literature reveals two dimensions of accountability: forward-looking responsibility, which focuses on identifying stakeholders and their roles, and backwards-looking responsibility, which involves acknowledging the outcomes of LA systems. Overall, addressing accountability is a significant challenge for HEIs. The accountability principle also overlaps with principles such as Safety, Security and Transparency. In their work, \cite{Reidenberg_Schaub_2018} suggest that implementing audit measures (mechanisms to inspect what, how, and why predictions and automated decisions were made) could not only improve algorithmic explainability but also support who (developers, admin staff, users) can be held accountable. In summary, the reviewed papers did not provide clear guidelines regarding the allocation of responsibilities among roles or the specific actions to be taken (e.g., escalation procedures) if harm arises from the use of LA systems.

\subsubsection{Safety}
The Alan Turing Institute's guidelines for safe AI systems \cite{Lesli-alan-turing_2019} highlight accuracy, reliability, and robustness as essential technical characteristics necessary to ensure AI functions safely and avoids harmful outcomes. In the context of LA, components such as data, decision algorithms, and applications (e.g., dashboards, alert systems) must be designed, deployed, and monitored to minimise errors, ensure consistent behaviour aligned with LA goals, and produce trustworthy predictions. To enhance safety it is crucial to provide end users—students, academic staff, and administrators—with clear documentation outlining the responsible use of LA systems, including guidance on interpreting data and engaging with at-risk students \cite{Howell_Roberts_Seaman_2018}. Developers should also receive clear conceptual frameworks and usage guidelines for LA systems \cite{Tabassi-NIST_2023}. Reliability in decision algorithms is critical; thus, establishing measurable goals for accuracy and expected model performance is vital. Specific considerations regarding acceptable error rates and performance metrics should be implemented. It is important to acknowledge that various factors—such as the choice of machine learning algorithms, missing data, and data noise—can influence predictions. Therefore, setting checkpoints for training and testing data is recommended \cite{microsoft_2022}. Additionally, ensuring data accuracy is paramount; research by \cite{Prinsloo_Slade_2013} underscores the need for HEI policies that guarantee access to up-to-date student data to prevent unreliable predictions. Despite these recommendations, there seems to still be a big gap in the literature on methods and actions that HEIs could put into practice to ensure that LA systems minimise harm to staff and students, whether psychological, emotional or academic. 

\subsubsection{Security}

The security principle includes the implementation of technical, administrative and physical controls to mitigate risks and prevent information assets from being accidentally or deliberately compromised. Key aspects\footnote{GDPR Article 32:1b: Security measures must \textit{"ensure the ongoing confidentiality, integrity, availability and resilience of processing systems and services"}} include ensuring confidentiality by controlling access to sensitive data about students and staff, maintaining integrity by preventing unauthorised data alteration or deletion, and guaranteeing availability for authorised users to access LA systems promptly \cite{Karunaratne_2021}.
Literature around Security indicates a dual focus on privacy and data security concerns \cite{Eleni_2023, Cormack_2016, Steiner_Kickmeier-Rust_Albert_2016};  
and both principles are strongly interlinked. For instance, implementing strong security measures (like data anonymisation) is crucial for protecting personal information, thus supporting privacy. Similarly, adhering to data privacy regulations comprises technical measures and policies to restrict unauthorized access or disclosure of personal information. 
Given that educational institutions collect extensive socio-demographic and progress data from students, any breach of this information could have detrimental effects on individuals and institutions alike. 
Therefore, the acquisition, processing, storage, and disposal of personal and sensitive data must adhere to strict legal and regulatory compliance standards \cite{Karunaratne_2021}. 
In this context, \cite{Skene_Winer_Kustra_2022} presents a compilation of ethical data governance considerations, which encompass data security aspects such as data process (public, sensitive, personal, high-risk), data storage (local vs. remote), and data audit plans. In the same line, the work by \cite{Drachsler_Greller_2016} compiles reference questions for managers and decision-makers to consider when implementing LA data security. 
Other researchers, however, who have looked at security in LA, have used the General Data Protection Regulation (GDPR) as the main guidance to protect personal data; for example, \cite{Cormack_2016} and \cite{Steiner_Kickmeier-Rust_Albert_2016} propose data security aligned with GDPR. 
While much of the existing literature emphasises legal frameworks and identifies data security issues, there remains a gap in specific operational and technical controls, which can hinder HEIs from effectively delivering the security principle within their LA solutions. Security controls should be tailored to align with organisational objectives and technological capabilities and evolve in response to advancements in technology, the latter being a great challenge for HEIs. 


\subsubsection{Fairness and bias} 

The principles of fairness and bias have garnered considerable attention within the Learning Analytics (LA) community, exemplified by dedicated workshops such as FairLAK\footnote{\url{https://sites.google.com/view/fairlak}}. Research has underscored the necessity of evaluating LA algorithms for biases and implementing effective mitigation strategies \cite{rets2023six}. Some studies focus on fairness metrics to assess biases affecting specific groups, including minority ethnic students \cite{bayer2021learning}. Advancing this work, Deho and colleagues \cite{deho2022existing} shifted from merely detecting bias to actively mitigating it, conducting a comparative evaluation of selected bias mitigation approaches. Their findings reveal that fairness lacks a universal definition, making the choice of definition a crucial first step in determining appropriate mitigation strategies. Furthermore, both studies indicate that enhancing fairness in LA systems may not need a compromise on predictive performance.
\cite{rets2023six} recommend the de-weighting or removal of sensitive attributes (and potential proxies, such as socio-economic status) from the training data of LA algorithms. However, Deho et al. \cite{deho2022should} clarify that the inclusion or exclusion of a protected attribute impacts performance and fairness only if it is correlated with the target label and deemed significant. Importantly, LA models that demonstrate fairness based on historical data may not maintain this fairness when applied to current or future datasets. Deho and colleagues therefore advocate for ensuring robustness against dataset drifts prior to deployment \cite{deho2024past}.
In a recent survey on biases in education, Li et al. \cite{li2023moral} emphasised the importance of considering intersectionality—how multiple sensitive attributes like gender and ethnicity interact—when evaluating algorithmic bias. They cautioned that applying fairness metrics to inappropriate tasks could lead to false conclusions and potentially harmful decisions. On the social aspect of fairness, \cite{tsai2021more} utilised questionnaires to gather insights from students and staff regarding the implications of bias in decision-making processes. Students expressed concerns about bias perpetuation and the fear of unfair assessments, while staff highlighted apprehensions regarding decisions made about them based on LA, such as managers using LA for performance evaluations.
While the existing identification and mitigation methods for bias provide valuable insights for HEIs, there remains a significant gap. The lack of clear guidelines on which definitions of fairness should be adopted based on specific objectives, as well as which bias mitigation methods are most suitable depending on the context (data, algorithm, etc.), leaves many open questions when operationalising this principle. 



\subsubsection{Transparency} 
The principle of transparency is crucial in ensuring that all stakeholders involved in a LA system (students, staff, and other relevant parties) are well-informed about its operations. Numerous studies highlight the necessity of transparency in LA \cite{Mahrishi_Abbas_Siddiqui_2024, Tzimas_Demetriadis_2021, Cerratto-Pargman_McGrath_2021}, yet many fall short of providing practical guidance on how to achieve it. Drachsler and Greller \cite{Drachsler_Greller_2016} note the inherent complexity in data collection and algorithmic processes, emphasising the challenge of conveying this information to non-technical stakeholders, including learners, teachers, and education managers. They advocate for giving data subjects access to their analytics results, empowering them to decide whether to seek pedagogical support or interventions, thereby placing the learner in control. The work also stresses the importance of obtaining clear consent prior to data collection, including the need for straightforward yes/no questions and the option to opt-out without repercussions. Hakami \cite{hakami2020learning} reviewed 37 LAK papers mentioning LA and Transparency. They highlighted the need to ensure stakeholders know how the algorithm works (algorithmic transparency), the need to set policies that reveal what data is collected and how they are used (institutional transparency), and the need to inform learners that they are being tracked (transparency and data). They also highlight that transparency can enhance understanding, sense-making and reflection, technology acceptance and adoption and trust. Tsai et al.\cite{tsai2020privacy} further highlight the significance of effective communication in achieving transparency, while Veljanova et al. \cite{veljanova2023operationalising} propose technical design features aimed at operationalising transparency within LA systems. Collectively, these works demonstrate the multifaceted nature of transparency, illustrating how open communication regarding data practices, algorithms, and decision-making processes fosters a more informed environment for all users involved in LA. However, significant gaps remain regarding when and how these communications should be implemented, as well as the most effective strategies to ensure that both students and staff receive, understand, and assimilate this information. \cite{tsai2020privacy} suggest practical steps such as organising workshops or meetings with students and incorporating relevant training on digital literacy into academic development programmes. These initiatives aim to raise awareness about the importance of data protection and empower students to take informed actions regarding their data in the context of LA. However, despite these valuable recommendations, the effectiveness of these proposed actions has yet to be evaluated.

\subsubsection{Privacy}

A wide range of studies have considered the privacy implications of LA. In a concept mapping exercise with experts, privacy as well as transparency were identified as the most important elements of LA policy \cite{Scheffel_Tsai_Gasevic_Drachsler_2019}. A key issue identified in prior studies is how and whether the student has agency over use of their data in the LA system. Recommendations include giving students the option to opt-out \cite{Tsai_Gasevic_2017}. Alternatively, LA could be offered on an opt-in basis \cite{prinsloo_2022} in which LA is presented in terms of how it can improve their learning experience \cite{Gosch_Andrews_Barreiros_Leitner_Staudegger_Ebner_Lindstaedt_2021}. To ensure consent is genuinely informed \cite{West_Huijser_Heath_2016, Tsai_Gasevic_2017, Patterson_York_Maxham_Molina_Mabrey_2023} students need greater awareness of what data is used and how \cite{prinsloo_2022} and the expected impact of granting or withdrawing consent \cite{Alzahrani_Tsai_Aljohani_Whitelock-wainwright_Gasevic_2023}. Workshops or meetings with students may be used to ensure students have appropriate knowledge of data literacy and data protection \cite{Tsai_Whitelock-Wainwright_=2020}. Consent-seeking procedures should be defined at an early stage \cite{Alzahrani_Tsai_Iqbal_2023} to help ensure initial or changing student preferences can be handled in the LA infrastructure \cite{Eleni_2023}. Similarly, staff should be given better guidance on the appropriate use of data and also the consequences of misuse (e.g. loss of confidentiality, negative publicity, legal action) \cite{Francis_Avoseh_Card_2023}. Such guidance could be informed by a Privacy Impact Assessment, covering impact on individuals, groups and wider society \cite{Clarke_2018}. Privacy should be initially considered at the point of the initial business case as part of the risk analysis for the initiative \cite{Clarke_2018}. Data should be anonymised wherever possible \cite{Tsai_Gasevic_2017, Tsai_Whitelock-Wainwright_=2020}, for example when aggregated to inform curriculum improvements \cite{Gosch_Andrews_Barreiros_Leitner_Staudegger_Ebner_Lindstaedt_2021}. More broadly, data governance guidelines should inform data sharing and ownership \cite{Alzahrani_Tsai_Aljohani_Whitelock-wainwright_Gasevic_2023} and be used to continually assess data access rights for different stakeholders \cite{Alzahrani_Tsai_Iqbal_2023}, minimising access to student data \cite{Tsai_Whitelock-Wainwright_=2020}. Privacy enhancing technologies such as anonymisation, encryption and digital signatures should be considered to improve the security of personal data \cite{prinsloo_2022} and any stakeholders should have access to an independent complaints body if they have any grievance over how their data has been accessed and used \cite{Clarke_2018}.

\subsubsection{Explainability}

The principle of explainability highlights the need for LA systems to offer clear insights into how their predictions and decisions are made. For example, \cite{Lunich_Keller_2024} note that traditional machine learning methods, like decision trees, provide higher interpretability compared to modern deep learning models, which often function as "black boxes." As a result, despite their lower accuracy, these interpretable systems may be preferred in scenarios where understanding the decision-making process is critical. This observation is further supported by Gunasekara et al. \cite{gunasekara2024explainability}, who reviewed explainability research within Educational Data Mining (EDM) and Learning Analytics, underscoring the importance of clarity in these systems. To address the explainability of more complex models, Li et al. \cite{li2024trustworthy} utilise two widely used explainable AI tools: Local Interpretable Model-agnostic Explanations (LIME) and Shapley Additive Explanations (SHAP). These tools, also adopted by well-known predictive analytics platforms \cite{hlosta2022predictive}, help break down complex model outputs into understandable components. Li and colleagues stress the importance of developing robust evaluation metrics to assess the quality of these explanations. Indeed, Gunasekara et al.'s review \cite{gunasekara2024explainability} reveals a critical gap in research regarding the metrics necessary to effectively evaluate the quality and utility of these explanations.
Taken together, these works reflect the complex nature of explainability in LA. While existing studies underscore the importance of offering transparent and comprehensible explanations for decision-making processes, a significant gap remains in providing practical guidance on how to implement explainability effectively in educational settings. The challenge lies not only in ensuring that explanations are available but also in communicating them in ways that are meaningful and accessible to a diverse range of users, including both students and educators.

\section{Responsible AI Framework for Learning Analytics}\label{sec:rai-framework}

\begin{figure}[t!]
  \centering
  \includegraphics[scale=0.25, angle=90]{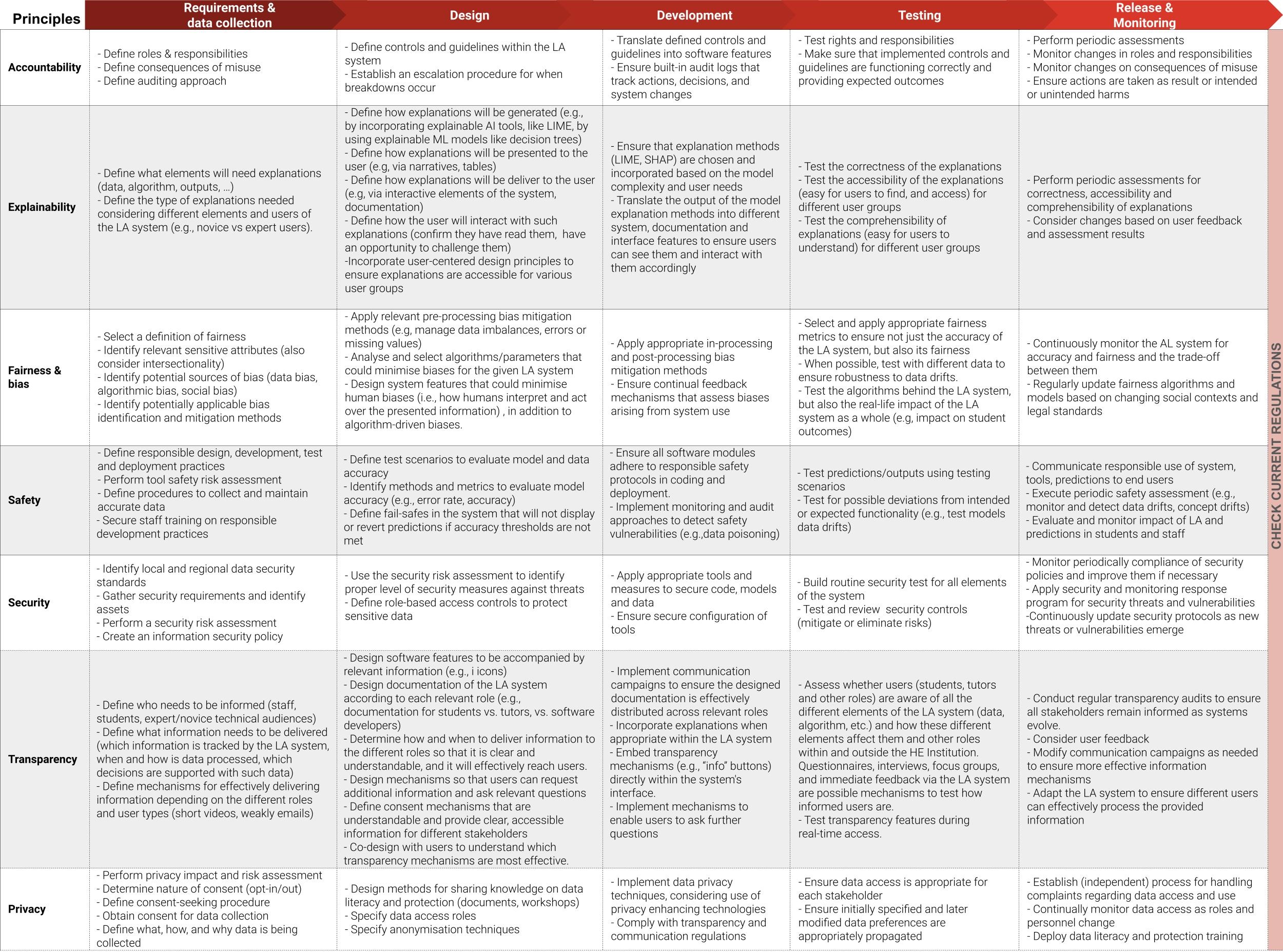}
  \caption{Responsible AI Framework for Learning Analytics in HEI}
  \label{fig:rai-framework-proposal}
\end{figure}

In this section, we introduce our Responsible AI (RAI) framework tailored to Learning Analytics (LA) in Higher Education (HE) - see Figure \ref{fig:rai-framework-proposal}. The primary aim of this framework is to provide higher education institutions with actionable guidance on how to incorporate responsible AI principles effectively into their LA initiatives. Recognising that institutions are at various stages of their LA adoption, we have structured the framework to follow the stages of the software development lifecycle: Requirements and Data Collection, Design, Development, Testing, Release, and Monitoring. By aligning the framework with these stages, we address a key limitation of many existing resources, allowing HEIs to engage with the specific stage of development they are currently in. This approach enables a more flexible, actionable pathway for integrating responsible AI principles.

While our ultimate goal is to provide both a list of actions HEIs can take to ensure their LA systems incorporate responsible AI principles and how to implement these actions, our literature review reveals a significant lack of real-world examples of how HEIs have operationalised these principles—if they have done so at all. We acknowledge that this leaves our framework incomplete, particularly in offering specific, practical steps that have already been tested in the field. However, we see this as an opportunity for continued growth. Our ambition is to refine this resource in collaboration with the wider academic and practitioner community, learning from best practices as they emerge.

For the purposes of this paper, a version of our proposed framework is summarised in Figure \ref{fig:rai-framework-proposal} and accessible via an anonymised URL\footnote{Project repository:\url{https://osf.io/at97f/?view_only=15b10f42abaa466691ffcce8a61226c1}}, where we have presented the relevant elements through a PowerPoint presentation (due to the constraints of the double-blind review process). The framework is hosted on a dedicated project website, where each step is linked to available documentation and real-world case studies from HEIs. This evolving resource will allow the community to contribute relevant materials, such as code libraries, consent forms, and other practical examples, fostering a collaborative environment where institutions can learn from one another.

Ultimately, we hope that this framework becomes a dynamic tool for HEIs seeking to responsibly implement LA systems, enabling them to align their practices with responsible AI principles.

\section{Discussion and Conclusions}\label{sec:conclusion}
In this paper, we have sought to address a pressing need within Higher Education Institutions (HEIs) for practical guidance on implementing Responsible AI principles within Learning Analytics solutions. Our aim is to establish a comprehensive Responsible AI framework tailored specifically for LA applications in HE. This framework is rooted in existing literature, but we have taken an important step further by examining not only high-level principles but also detailed accounts from studies that have operationalised these principles in various contexts. We have gathered lessons learned and challenges encountered, providing a richer foundation for our proposed framework.

Our framework is organised around the Software Development Lifecycle, encompassing critical stages from requirements gathering and data collection to deployment and monitoring. By doing this, we aim to create a structured approach that allows HEIs to systematically integrate Responsible AI principles into their LA practices. Each stage of the lifecycle includes a list of actionable items based on insights gleaned from the literature, offering guidance on operationalising these principles.

This resource, which is accessible online (URL withheld due to the double-blind review process), is intentionally a work in progress. We envision it as a starting point for collaboration and dialogue between the academic community and practitioners. Discussing our framework at relevant venues is crucial for refining and enhancing this resource, as it enables us to gather feedback and share best practices with stakeholders in the LA community.

We acknowledge that there may be relevant works from other disciplines—such as Computer Science and Sociology—that explore aspects of Responsible AI, albeit not specifically within the HE context. While these works have not been incorporated into this initial literature overview, our goal is to continually expand our understanding and enrich our framework with interdisciplinary insights that can aid HEIs in operationalising Responsible AI principles.

It is important to note that for certain principles, particularly accountability, there remain significant gaps in practical examples. We hope that this paper will stimulate new research directions that can help the community address these challenges. By highlighting these gaps, we aim to foster a collaborative environment where researchers can share insights and develop concrete strategies for accountability in LA systems.

In summary, our proposed Responsible AI framework serves as a vital resource for HEIs looking to implement ethical and responsible practices in their learning analytics efforts. We believe that by engaging with the broader academic and practitioner communities, we can collectively enhance the application of Responsible AI principles, ultimately benefiting both educators and students in the evolving landscape of higher education.

\begin{acks}
This project has received funding from the UK Government - Department for Science, Innovation and Technology (DSIT). This funding is administered by UKRI.
\end{acks}

\bibliographystyle{ACM-Reference-Format}
\bibliography{bibliography}


\begin{thebibliography}{63}


\ifx \showCODEN    \undefined \def \showCODEN     #1{\unskip}     \fi
\ifx \showDOI      \undefined \def \showDOI       #1{#1}\fi
\ifx \showISBNx    \undefined \def \showISBNx     #1{\unskip}     \fi
\ifx \showISBNxiii \undefined \def \showISBNxiii  #1{\unskip}     \fi
\ifx \showISSN     \undefined \def \showISSN      #1{\unskip}     \fi
\ifx \showLCCN     \undefined \def \showLCCN      #1{\unskip}     \fi
\ifx \shownote     \undefined \def \shownote      #1{#1}          \fi
\ifx \showarticletitle \undefined \def \showarticletitle #1{#1}   \fi
\ifx \showURL      \undefined \def \showURL       {\relax}        \fi
\providecommand\bibfield[2]{#2}
\providecommand\bibinfo[2]{#2}
\providecommand\natexlab[1]{#1}
\providecommand\showeprint[2][]{arXiv:#2}

\bibitem[Alliance(2022)]%
        {Circleu_2022}
\bibfield{author}{\bibinfo{person}{Circle U. European~University Alliance}.} \bibinfo{year}{2022}\natexlab{}.
\newblock \bibinfo{title}{Legal and Ethical Aspects of Using Learning Analytics from a Univeristy Perspective}.
\newblock
\newblock


\bibitem[Alvarez et~al\mbox{.}(2024)]%
        {alvarez2024policy}
\bibfield{author}{\bibinfo{person}{Jose~M Alvarez}, \bibinfo{person}{Alejandra~Bringas Colmenarejo}, \bibinfo{person}{Alaa Elobaid}, \bibinfo{person}{Simone Fabbrizzi}, \bibinfo{person}{Miriam Fahimi}, \bibinfo{person}{Antonio Ferrara}, \bibinfo{person}{Siamak Ghodsi}, \bibinfo{person}{Carlos Mougan}, \bibinfo{person}{Ioanna Papageorgiou}, \bibinfo{person}{Paula Reyero}, {et~al\mbox{.}}} \bibinfo{year}{2024}\natexlab{}.
\newblock \showarticletitle{Policy advice and best practices on bias and fairness in AI}.
\newblock \bibinfo{journal}{\emph{Ethics and Information Technology}} (\bibinfo{year}{2024}).
\newblock


\bibitem[Alzahrani et~al\mbox{.}(2023a)]%
        {Alzahrani_Tsai_Aljohani_Whitelock-wainwright_Gasevic_2023}
\bibfield{author}{\bibinfo{person}{Asma~Shannan Alzahrani}, \bibinfo{person}{Yi-Shan Tsai}, \bibinfo{person}{Naif Aljohani}, \bibinfo{person}{Emma Whitelock-wainwright}, {and} \bibinfo{person}{Dragan Gasevic}.} \bibinfo{year}{2023}\natexlab{a}.
\newblock \showarticletitle{Do teaching staff trust stakeholders and tools in learning analytics? A mixed methods study}.
\newblock \bibinfo{journal}{\emph{Educational technology research and development}} (\bibinfo{date}{Aug.} \bibinfo{year}{2023}).
\newblock
\showISSN{1042-1629, 1556-6501}


\bibitem[Alzahrani et~al\mbox{.}(2023b)]%
        {Alzahrani_Tsai_Iqbal_2023}
\bibfield{author}{\bibinfo{person}{Asma~Shannan Alzahrani}, \bibinfo{person}{Yi-Shan Tsai}, \bibinfo{person}{Sehrish Iqbal}, \bibinfo{person}{Pedro Manuel~Moreno Marcos}, \bibinfo{person}{Maren Scheffel}, \bibinfo{person}{Hendrik Drachsler}, \bibinfo{person}{Carlos~Delgado Kloos}, \bibinfo{person}{Naif Aljohani}, {and} \bibinfo{person}{Dragan Gasevic}.} \bibinfo{year}{2023}\natexlab{b}.
\newblock \showarticletitle{Untangling Connections between Challenges in the Adoption of Learning Analytics in Higher Education}.
\newblock \bibinfo{journal}{\emph{Education and Information Technologies}} (\bibinfo{date}{April} \bibinfo{year}{2023}).
\newblock
\showISSN{1573-7608}


\bibitem[Amazon({[n.\,d.]})]%
        {amazon_na}
\bibfield{author}{\bibinfo{person}{Amazon}.} \bibinfo{year}{[n.\,d.]}\natexlab{}.
\newblock \bibinfo{title}{Responsible AI}.
\newblock
\newblock
\urldef\tempurl%
\url{https://aws.amazon.com/machine-learning/responsible-ai/}
\showURL{%
\tempurl}


\bibitem[Barletta et~al\mbox{.}(2023)]%
        {barletta2023rapid}
\bibfield{author}{\bibinfo{person}{Vita~Santa Barletta}, \bibinfo{person}{Danilo Caivano}, \bibinfo{person}{Domenico Gigante}, {and} \bibinfo{person}{Azzurra Ragone}.} \bibinfo{year}{2023}\natexlab{}.
\newblock \showarticletitle{A rapid review of responsible ai frameworks: How to guide the development of ethical ai}. In \bibinfo{booktitle}{\emph{Proceedings of the 27th International Conference on Evaluation and Assessment in Software Engineering}}.
\newblock


\bibitem[Bayer et~al\mbox{.}(2021)]%
        {bayer2021learning}
\bibfield{author}{\bibinfo{person}{Vaclav Bayer}, \bibinfo{person}{Martin Hlosta}, {and} \bibinfo{person}{Miriam Fernandez}.} \bibinfo{year}{2021}\natexlab{}.
\newblock \showarticletitle{Learning analytics and fairness: do existing algorithms serve everyone equally?}. In \bibinfo{booktitle}{\emph{International Conference on Artificial Intelligence in Education}}. Springer.
\newblock


\bibitem[Cerratto~Pargman and McGrath(2021)]%
        {Cerratto-Pargman_McGrath_2021}
\bibfield{author}{\bibinfo{person}{Teresa Cerratto~Pargman} {and} \bibinfo{person}{Cormac McGrath}.} \bibinfo{year}{2021}\natexlab{}.
\newblock \showarticletitle{Mapping the Ethics of Learning Analytics in Higher Education: A Systematic Literature Review of Empirical Research}.
\newblock \bibinfo{journal}{\emph{Journal of Learning Analytics}} \bibinfo{number}{2} (\bibinfo{date}{Sept.} \bibinfo{year}{2021}).
\newblock
\showISSN{19297750}
\urldef\tempurl%
\url{https://doi.org/10.18608/jla.2021.1}
\showDOI{\tempurl}


\bibitem[Clarke(2018)]%
        {Clarke_2018}
\bibfield{author}{\bibinfo{person}{Roger Clarke}.} \bibinfo{year}{2018}\natexlab{}.
\newblock \showarticletitle{Guidelines for the responsible application of data analytics}.
\newblock \bibinfo{journal}{\emph{Computer Law \& Security Review}} (\bibinfo{date}{June} \bibinfo{year}{2018}).
\newblock
\showISSN{02673649}


\bibitem[Cormack(2016)]%
        {Cormack_2016}
\bibfield{author}{\bibinfo{person}{Andrew~Nicholas Cormack}.} \bibinfo{year}{2016}\natexlab{}.
\newblock \showarticletitle{A Data Protection Framework for Learning Analytics}.
\newblock  (\bibinfo{date}{April} \bibinfo{year}{2016}).
\newblock
\showISSN{1929-7750}
\urldef\tempurl%
\url{https://doi.org/10.18608/jla.2016.31.6}
\showDOI{\tempurl}


\bibitem[Corrin et~al\mbox{.}(2019)]%
        {ethics-australian-hei_2019}
\bibfield{author}{\bibinfo{person}{Linda Corrin}, \bibinfo{person}{Gregor Kennedy}, \bibinfo{person}{Sarah French}, \bibinfo{person}{Simon Buckingham~Shum}, \bibinfo{person}{Kirsty Kitto}, \bibinfo{person}{Abelardo Pardo}, \bibinfo{person}{Deborah West}, \bibinfo{person}{Negin Mirriahi}, {and} \bibinfo{person}{Cassandra Colvin}.} \bibinfo{year}{2019}\natexlab{}.
\newblock \bibinfo{title}{The Ethics of Learning Analytics in Australian Higher Education. A Discussion Paper}.
\newblock
\newblock
\urldef\tempurl%
\url{https://resolver.ebscohost.com/Redirect/PRL?EPPackageLocationID=2418090.1356208.27427588&epcustomerid=s2947694}
\showURL{%
\tempurl}


\bibitem[Crompton and Burke(2023)]%
        {Crompton_Burke_2023}
\bibfield{author}{\bibinfo{person}{Helen Crompton} {and} \bibinfo{person}{Diane Burke}.} \bibinfo{year}{2023}\natexlab{}.
\newblock \showarticletitle{Artificial intelligence in higher education: the state of the field}.
\newblock \bibinfo{journal}{\emph{International Journal of Educational Technology in Higher Education}} (\bibinfo{date}{April} \bibinfo{year}{2023}).
\newblock
\showISSN{2365-9440}
\urldef\tempurl%
\url{https://doi.org/10.1186/s41239-023-00392-8}
\showDOI{\tempurl}


\bibitem[Deho et~al\mbox{.}(2022a)]%
        {deho2022should}
\bibfield{author}{\bibinfo{person}{Oscar~Blessed Deho}, \bibinfo{person}{Srecko Joksimovic}, \bibinfo{person}{Jiuyong Li}, \bibinfo{person}{Chen Zhan}, \bibinfo{person}{Jixue Liu}, {and} \bibinfo{person}{Lin Liu}.} \bibinfo{year}{2022}\natexlab{a}.
\newblock \showarticletitle{Should learning analytics models include sensitive attributes? Explaining the why}.
\newblock \bibinfo{journal}{\emph{IEEE Transactions on Learning Technologies}} \bibinfo{volume}{16}, \bibinfo{number}{4} (\bibinfo{year}{2022}), \bibinfo{pages}{560--572}.
\newblock


\bibitem[Deho et~al\mbox{.}(2024)]%
        {deho2024past}
\bibfield{author}{\bibinfo{person}{Oscar~Blessed Deho}, \bibinfo{person}{Lin Liu}, \bibinfo{person}{Jiuyong Li}, \bibinfo{person}{Jixue Liu}, \bibinfo{person}{Chen Zhan}, {and} \bibinfo{person}{Srecko Joksimovic}.} \bibinfo{year}{2024}\natexlab{}.
\newblock \showarticletitle{When the past!= the future: Assessing the Impact of Dataset Drift on the Fairness of Learning Analytics Models}.
\newblock \bibinfo{journal}{\emph{IEEE Transactions on Learning Technologies}} (\bibinfo{year}{2024}).
\newblock


\bibitem[Deho et~al\mbox{.}(2022b)]%
        {deho2022existing}
\bibfield{author}{\bibinfo{person}{Oscar~Blessed Deho}, \bibinfo{person}{Chen Zhan}, \bibinfo{person}{Jiuyong Li}, \bibinfo{person}{Jixue Liu}, \bibinfo{person}{Lin Liu}, {and} \bibinfo{person}{Thuc Duy~Le}.} \bibinfo{year}{2022}\natexlab{b}.
\newblock \showarticletitle{How do the existing fairness metrics and unfairness mitigation algorithms contribute to ethical learning analytics?}
\newblock \bibinfo{journal}{\emph{British Journal of Educational Technology}} (\bibinfo{year}{2022}).
\newblock


\bibitem[Drachsler and Greller(2016)]%
        {Drachsler_Greller_2016}
\bibfield{author}{\bibinfo{person}{Hendrik Drachsler} {and} \bibinfo{person}{Wolfgang Greller}.} \bibinfo{year}{2016}\natexlab{}.
\newblock \showarticletitle{Privacy and Analytics – it’s a DELICATE Issue A Checklist for Trusted Learning Analytics}. \bibinfo{address}{Edinburgh}.
\newblock


\bibitem[Eleni(2023)]%
        {Eleni_2023}
\bibfield{author}{\bibinfo{person}{Polydorou Eleni}.} \bibinfo{year}{2023}\natexlab{}.
\newblock \bibinfo{booktitle}{\emph{Towards a Secure and Privacy Compliant Framework for Educational Data Mining}}.
\newblock \bibinfo{publisher}{Springer Nature Switzerland}.
\newblock
\showISBNx{978-3-031-33079-7}


\bibitem[Ferguson(2019)]%
        {ferguson2019ethical}
\bibfield{author}{\bibinfo{person}{Rebecca Ferguson}.} \bibinfo{year}{2019}\natexlab{}.
\newblock \showarticletitle{Ethical Challenges for Learning Analytics.}
\newblock \bibinfo{journal}{\emph{Journal of Learning Analytics}} (\bibinfo{year}{2019}).
\newblock


\bibitem[for Ethical Al~in Education(2021)]%
        {ieaie_2021}
\bibfield{author}{\bibinfo{person}{The~Institute for Ethical Al~in Education}.} \bibinfo{year}{2021}\natexlab{}.
\newblock \bibinfo{booktitle}{\emph{The Ethical Framework for AI in Education}}.
\newblock
\urldef\tempurl%
\url{https://www.buckingham.ac.uk/research/research-in-applied-computing/the-institute-for-ethical-ai-in-education/}
\showURL{%
\tempurl}


\bibitem[Francis et~al\mbox{.}(2023)]%
        {Francis_Avoseh_Card_2023}
\bibfield{author}{\bibinfo{person}{Mary Francis}, \bibinfo{person}{Mejai Bola~Mike Avoseh}, \bibinfo{person}{Karen Card}, \bibinfo{person}{Lisa Newland}, {and} \bibinfo{person}{Kevin Streff}.} \bibinfo{year}{2023}\natexlab{}.
\newblock \showarticletitle{Student Privacy and Learning Analytics: Investigating the Application of Privacy within a Student Success Information System in Higher Education}.
\newblock \bibinfo{journal}{\emph{Journal of Learning Analytics}} (\bibinfo{year}{2023}).
\newblock
\showISSN{1929-7750}


\bibitem[Google(2023)]%
        {Google_2023}
\bibfield{author}{\bibinfo{person}{Google}.} \bibinfo{year}{2023}\natexlab{}.
\newblock \bibinfo{title}{AI Principles Progress Update 2023}.
\newblock
\newblock
\urldef\tempurl%
\url{https://ai.google/static/documents/ai-principles-2023-progress-update.pdf}
\showURL{%
\tempurl}


\bibitem[Gosch et~al\mbox{.}(2021)]%
        {Gosch_Andrews_Barreiros_Leitner_Staudegger_Ebner_Lindstaedt_2021}
\bibfield{author}{\bibinfo{person}{Nicole Gosch}, \bibinfo{person}{David Andrews}, \bibinfo{person}{Carla Barreiros}, \bibinfo{person}{Philipp Leitner}, \bibinfo{person}{Elisabeth Staudegger}, \bibinfo{person}{Martin Ebner}, {and} \bibinfo{person}{Stefanie Lindstaedt}.} \bibinfo{year}{2021}\natexlab{}.
\newblock \showarticletitle{Learning Analytics as a Service for Empowered Learners: From Data Subjects to Controllers}. In \bibinfo{booktitle}{\emph{LAK21: 11th International Learning Analytics and Knowledge Conference}}.
\newblock
\showISBNx{978-1-4503-8935-8}
\urldef\tempurl%
\url{https://doi.org/10.1145/3448139.3448186}
\showDOI{\tempurl}


\bibitem[Gunasekara and Saarela(2024)]%
        {gunasekara2024explainability}
\bibfield{author}{\bibinfo{person}{Sachini Gunasekara} {and} \bibinfo{person}{Mirka Saarela}.} \bibinfo{year}{2024}\natexlab{}.
\newblock \showarticletitle{Explainability in Educational Data Mining and Learning Analytics: An Umbrella Review}. In \bibinfo{booktitle}{\emph{International conference on educational data mining}}. International Educational Data Mining Society.
\newblock


\bibitem[Hakami and Hern{\'a}ndez~Leo(2020)]%
        {hakami2020learning}
\bibfield{author}{\bibinfo{person}{Eyad Hakami} {and} \bibinfo{person}{Davinia Hern{\'a}ndez~Leo}.} \bibinfo{year}{2020}\natexlab{}.
\newblock \showarticletitle{How are learning analytics considering the societal values of fairness, accountability, transparency and human well-being?: A literature review}.
\newblock \bibinfo{journal}{\emph{Mart{\'\i}nez-Mon{\'e}s A, {\'A}lvarez A, Caeiro-Rodr{\'\i}guez M, Dimitriadis Y, editors. LASI-SPAIN 2020: Learning Analytics Summer Institute Spain 2020: Learning Analytics. Time for Adoption?; 2020 Jun 15-16; Valladolid, Spain. Aachen: CEUR; 2020.}} (\bibinfo{year}{2020}).
\newblock


\bibitem[Hlosta et~al\mbox{.}(2022)]%
        {hlosta2022predictive}
\bibfield{author}{\bibinfo{person}{Martin Hlosta}, \bibinfo{person}{Christothea Herodotou}, \bibinfo{person}{Tina Papathoma}, \bibinfo{person}{Anna Gillespie}, {and} \bibinfo{person}{Per Bergamin}.} \bibinfo{year}{2022}\natexlab{}.
\newblock \showarticletitle{Predictive learning analytics in online education: A deeper understanding through explaining algorithmic errors}.
\newblock \bibinfo{journal}{\emph{Computers and Education: Artificial Intelligence}} (\bibinfo{year}{2022}).
\newblock


\bibitem[Howell et~al\mbox{.}(2018)]%
        {Howell_Roberts_Seaman_2018}
\bibfield{author}{\bibinfo{person}{Joel~A. Howell}, \bibinfo{person}{Lynne~D. Roberts}, \bibinfo{person}{Kristen Seaman}, {and} \bibinfo{person}{David~C. Gibson}.} \bibinfo{year}{2018}\natexlab{}.
\newblock \showarticletitle{Are We on Our Way to Becoming a “Helicopter University”? Academics’ Views on Learning Analytics}.
\newblock \bibinfo{journal}{\emph{Technology, Knowledge and Learning}} (\bibinfo{date}{April} \bibinfo{year}{2018}).
\newblock
\showISSN{2211-1662, 2211-1670}
\urldef\tempurl%
\url{https://doi.org/10.1007/s10758-017-9329-9}
\showDOI{\tempurl}


\bibitem[IBM(2024)]%
        {IBM}
\bibfield{author}{\bibinfo{person}{IBM}.} \bibinfo{year}{2024}\natexlab{}.
\newblock \bibinfo{title}{AI ethics}.
\newblock
\newblock
\urldef\tempurl%
\url{https://www.ibm.com/impact/ai-ethics}
\showURL{%
\tempurl}


\bibitem[{JISC}(2023)]%
        {jisc_code_2023}
\bibfield{author}{\bibinfo{person}{{JISC}}.} \bibinfo{year}{2023}\natexlab{}.
\newblock \bibinfo{title}{Code of practice for learning analytics}.
\newblock
\newblock
\urldef\tempurl%
\url{https://repository.jisc.ac.uk/9204/1/code-of-practice-for-learning-analytics.pdf}
\showURL{%
\tempurl}


\bibitem[Karunaratne(2021)]%
        {Karunaratne_2021}
\bibfield{author}{\bibinfo{person}{Thashmee Karunaratne}.} \bibinfo{year}{2021}\natexlab{}.
\newblock \showarticletitle{For Learning Analytics to Be Sustainable under GDPR—Consequences and Way Forward}.
\newblock \bibinfo{journal}{\emph{Sustainability}} (\bibinfo{date}{Oct.} \bibinfo{year}{2021}).
\newblock
\showISSN{2071-1050}


\bibitem[Lesli(2019)]%
        {Lesli-alan-turing_2019}
\bibfield{author}{\bibinfo{person}{D Lesli}.} \bibinfo{year}{2019}\natexlab{}.
\newblock \bibinfo{title}{Understanding artificial intelligence ethics and safety: A guide for the responsible design and implementation of AI systems in the public sector}.
\newblock
\newblock
\urldef\tempurl%
\url{https://doi.org/10.5281/zenodo.3240529}
\showURL{%
\tempurl}


\bibitem[Li et~al\mbox{.}(2023)]%
        {li2023moral}
\bibfield{author}{\bibinfo{person}{Lin Li}, \bibinfo{person}{Lele Sha}, \bibinfo{person}{Yuheng Li}, \bibinfo{person}{Mladen Rakovi{\'c}}, \bibinfo{person}{Jia Rong}, \bibinfo{person}{Srecko Joksimovic}, \bibinfo{person}{Neil Selwyn}, \bibinfo{person}{Dragan Ga{\v{s}}evi{\'c}}, {and} \bibinfo{person}{Guanliang Chen}.} \bibinfo{year}{2023}\natexlab{}.
\newblock \showarticletitle{Moral machines or tyranny of the majority? A systematic review on predictive bias in education}. In \bibinfo{booktitle}{\emph{LAK23: 13th International Learning Analytics and Knowledge conference}}.
\newblock


\bibitem[Li et~al\mbox{.}(2024)]%
        {li2024trustworthy}
\bibfield{author}{\bibinfo{person}{Min-Jia Li}, \bibinfo{person}{Shun-Ting Li}, \bibinfo{person}{Albert~CM Yang}, \bibinfo{person}{Anna~YQ Huang}, {and} \bibinfo{person}{Stephen~JH Yang}.} \bibinfo{year}{2024}\natexlab{}.
\newblock \showarticletitle{Trustworthy and Explainable AI for Learning Analytics.}. In \bibinfo{booktitle}{\emph{LAK Workshops}}. \bibinfo{pages}{3--12}.
\newblock


\bibitem[Lünich and Keller(2024)]%
        {Lunich_Keller_2024}
\bibfield{author}{\bibinfo{person}{Marco Lünich} {and} \bibinfo{person}{Birte Keller}.} \bibinfo{year}{2024}\natexlab{}.
\newblock \showarticletitle{Explainable Artificial Intelligence for Academic Performance Prediction. An Experimental Study on the Impact of Accuracy and Simplicity of Decision Trees on Causability and Fairness Perceptions}. In \bibinfo{booktitle}{\emph{Proceedings of the 2024 ACM Conference on Fairness, Accountability, and Transparency}} \emph{(\bibinfo{series}{FAccT ’24})}. \bibinfo{publisher}{Association for Computing Machinery}, \bibinfo{address}{New York, NY, USA}.
\newblock
\showISBNx{9798400704505}
\urldef\tempurl%
\url{https://doi.org/10.1145/3630106.3658953}
\showDOI{\tempurl}


\bibitem[Mahrishi et~al\mbox{.}(2024)]%
        {Mahrishi_Abbas_Siddiqui_2024}
\bibfield{author}{\bibinfo{person}{Mehul Mahrishi}, \bibinfo{person}{Asad Abbas}, {and} \bibinfo{person}{Mohammad~Khubeb Siddiqui}.} \bibinfo{year}{2024}\natexlab{}.
\newblock \showarticletitle{Global Initiatives Towards Regulatory Frameworks for Artificial Intelligence (AI) in Higher Education}.
\newblock  (\bibinfo{year}{2024}).
\newblock
\urldef\tempurl%
\url{https://doi.org/10.1145/3672462}
\showDOI{\tempurl}


\bibitem[Mathrani et~al\mbox{.}(2021)]%
        {mathrani2021perspectives}
\bibfield{author}{\bibinfo{person}{Anuradha Mathrani}, \bibinfo{person}{Teo Susnjak}, \bibinfo{person}{Gomathy Ramaswami}, {and} \bibinfo{person}{Andre Barczak}.} \bibinfo{year}{2021}\natexlab{}.
\newblock \showarticletitle{Perspectives on the challenges of generalizability, transparency and ethics in predictive learning analytics}.
\newblock \bibinfo{journal}{\emph{Computers and Education Open}} (\bibinfo{year}{2021}).
\newblock


\bibitem[Microsoft(2022)]%
        {microsoft_2022}
\bibfield{author}{\bibinfo{person}{Microsoft}.} \bibinfo{year}{2022}\natexlab{}.
\newblock
\newblock
\urldef\tempurl%
\url{https://blogs.microsoft.com/on-the-issues/2022/06/21/microsofts-framework-for-building-ai-systems-responsibly/}
\showURL{%
\tempurl}


\bibitem[Nguyen et~al\mbox{.}(2023)]%
        {nguyen2023ethical}
\bibfield{author}{\bibinfo{person}{Andy Nguyen}, \bibinfo{person}{Ha~Ngan Ngo}, \bibinfo{person}{Yvonne Hong}, \bibinfo{person}{Belle Dang}, {and} \bibinfo{person}{Bich-Phuong~Thi Nguyen}.} \bibinfo{year}{2023}\natexlab{}.
\newblock \showarticletitle{Ethical principles for artificial intelligence in education}.
\newblock \bibinfo{journal}{\emph{Education and Information Technologies}} (\bibinfo{year}{2023}).
\newblock


\bibitem[Ntoutsi et~al\mbox{.}(2020)]%
        {ntoutsi2020bias}
\bibfield{author}{\bibinfo{person}{Eirini Ntoutsi}, \bibinfo{person}{Pavlos Fafalios}, \bibinfo{person}{Ujwal Gadiraju}, \bibinfo{person}{Vasileios Iosifidis}, \bibinfo{person}{Wolfgang Nejdl}, \bibinfo{person}{Maria-Esther Vidal}, \bibinfo{person}{Salvatore Ruggieri}, \bibinfo{person}{Franco Turini}, \bibinfo{person}{Symeon Papadopoulos}, \bibinfo{person}{Emmanouil Krasanakis}, {et~al\mbox{.}}} \bibinfo{year}{2020}\natexlab{}.
\newblock \showarticletitle{Bias in data-driven artificial intelligence systems—An introductory survey}.
\newblock \bibinfo{journal}{\emph{Wiley Interdisciplinary Reviews: Data Mining and Knowledge Discovery}} (\bibinfo{year}{2020}).
\newblock


\bibitem[of~Edinburgh(2017)]%
        {UoE_2017}
\bibfield{author}{\bibinfo{person}{University of Edinburgh}.} \bibinfo{year}{2017}\natexlab{}.
\newblock \bibinfo{title}{Learning Analytics Principles and Purposes}.
\newblock
\newblock
\urldef\tempurl%
\url{https://www.ed.ac.uk/files/atoms/files/learninganalyticsprinciples.pdf}
\showURL{%
\tempurl}


\bibitem[Patterson et~al\mbox{.}(2023)]%
        {Patterson_York_Maxham_Molina_Mabrey_2023}
\bibfield{author}{\bibinfo{person}{Chris Patterson}, \bibinfo{person}{Emily York}, \bibinfo{person}{Danielle Maxham}, \bibinfo{person}{Rudy Molina}, {and} \bibinfo{person}{Paul Mabrey}.} \bibinfo{year}{2023}\natexlab{}.
\newblock \showarticletitle{Applying a Responsible Innovation Framework in Developing an Equitable Early Alert System: A Case Study}.
\newblock \bibinfo{journal}{\emph{Journal of Learning Analytics}} (\bibinfo{date}{March} \bibinfo{year}{2023}).
\newblock
\showISSN{1929-7750}
\urldef\tempurl%
\url{https://doi.org/10.18608/jla.2023.7795}
\showDOI{\tempurl}


\bibitem[Prinsloo and Slade(2013)]%
        {Prinsloo_Slade_2013}
\bibfield{author}{\bibinfo{person}{Paul Prinsloo} {and} \bibinfo{person}{Sharon Slade}.} \bibinfo{year}{2013}\natexlab{}.
\newblock \showarticletitle{An evaluation of policy frameworks for addressing ethical considerations in learning analytics}. \bibinfo{address}{Belgium}.
\newblock


\bibitem[Prinsloo et~al\mbox{.}(2022)]%
        {prinsloo_2022}
\bibfield{author}{\bibinfo{person}{Paul Prinsloo}, \bibinfo{person}{Sharon Slade}, {and} \bibinfo{person}{Mohammad Khalil}.} \bibinfo{year}{2022}\natexlab{}.
\newblock \showarticletitle{The answer is (not only) technological: Considering student data privacy in learning analytics}.
\newblock \bibinfo{journal}{\emph{British Journal of Educational Technology}} (\bibinfo{year}{2022}).
\newblock


\bibitem[Reidenberg and Schaub(2018)]%
        {Reidenberg_Schaub_2018}
\bibfield{author}{\bibinfo{person}{Joel~R. Reidenberg} {and} \bibinfo{person}{Florian Schaub}.} \bibinfo{year}{2018}\natexlab{}.
\newblock \showarticletitle{Achieving Big Data Privacy in Education}.
\newblock \bibinfo{journal}{\emph{Theory and Research in Education}} (\bibinfo{date}{Nov.} \bibinfo{year}{2018}).
\newblock
\showISSN{1477-8785}


\bibitem[Rets et~al\mbox{.}(2023)]%
        {rets2023six}
\bibfield{author}{\bibinfo{person}{Irina Rets}, \bibinfo{person}{Christothea Herodotou}, {and} \bibinfo{person}{Anna Gillespie}.} \bibinfo{year}{2023}\natexlab{}.
\newblock \showarticletitle{Six Practical Recommendations Enabling Ethical Use of Predictive Learning Analytics in Distance Education.}
\newblock \bibinfo{journal}{\emph{Journal of Learning Analytics}} (\bibinfo{year}{2023}).
\newblock


\bibitem[Roberts et~al\mbox{.}(2017)]%
        {roberts2017ethical}
\bibfield{author}{\bibinfo{person}{Lynne~D Roberts}, \bibinfo{person}{Vanessa Chang}, {and} \bibinfo{person}{David Gibson}.} \bibinfo{year}{2017}\natexlab{}.
\newblock \showarticletitle{Ethical considerations in adopting a university-and system-wide approach to data and learning analytics}.
\newblock \bibinfo{journal}{\emph{Big data and learning analytics in higher education: Current theory and practice}} (\bibinfo{year}{2017}).
\newblock


\bibitem[Roberts et~al\mbox{.}(2016)]%
        {roberts2016student}
\bibfield{author}{\bibinfo{person}{Lynne~D Roberts}, \bibinfo{person}{Joel~A Howell}, \bibinfo{person}{Kristen Seaman}, {and} \bibinfo{person}{David~C Gibson}.} \bibinfo{year}{2016}\natexlab{}.
\newblock \showarticletitle{Student attitudes toward learning analytics in higher education:“The fitbit version of the learning world”}.
\newblock \bibinfo{journal}{\emph{Frontiers in psychology}} (\bibinfo{year}{2016}).
\newblock


\bibitem[Romero and Ventura(2020)]%
        {romero2020educational}
\bibfield{author}{\bibinfo{person}{Cristobal Romero} {and} \bibinfo{person}{Sebastian Ventura}.} \bibinfo{year}{2020}\natexlab{}.
\newblock \showarticletitle{Educational data mining and learning analytics: An updated survey}.
\newblock \bibinfo{journal}{\emph{Wiley interdisciplinary reviews: Data mining and knowledge discovery}} (\bibinfo{year}{2020}).
\newblock


\bibitem[Scheffel et~al\mbox{.}(2019)]%
        {Scheffel_Tsai_Gasevic_Drachsler_2019}
\bibfield{author}{\bibinfo{person}{Maren Scheffel}, \bibinfo{person}{Yi-Shan Tsai}, \bibinfo{person}{Dragan Ga\v{s}evi\'{c}}, {and} \bibinfo{person}{Hendrik Drachsler}.} \bibinfo{year}{2019}\natexlab{}.
\newblock \showarticletitle{Policy Matters: Expert Recommendations for Learning Analytics Policy}. In \bibinfo{booktitle}{\emph{Transforming Learning with Meaningful Technologies: 14th European Conference on Technology Enhanced Learning, EC-TEL 2019}}.
\newblock


\bibitem[Sghir et~al\mbox{.}(2023)]%
        {sghir2023recent}
\bibfield{author}{\bibinfo{person}{Nabila Sghir}, \bibinfo{person}{Amina Adadi}, {and} \bibinfo{person}{Mohammed Lahmer}.} \bibinfo{year}{2023}\natexlab{}.
\newblock \showarticletitle{Recent advances in Predictive Learning Analytics: A decade systematic review (2012--2022)}.
\newblock \bibinfo{journal}{\emph{Education and information technologies}} (\bibinfo{year}{2023}).
\newblock


\bibitem[Skene et~al\mbox{.}(2022)]%
        {Skene_Winer_Kustra_2022}
\bibfield{author}{\bibinfo{person}{Allyson Skene}, \bibinfo{person}{Laura Winer}, {and} \bibinfo{person}{Erika Kustra}.} \bibinfo{year}{2022}\natexlab{}.
\newblock \showarticletitle{Clouds in the silver lining of learning analytics: ethical tensions for Educational Developers}.
\newblock \bibinfo{journal}{\emph{International Journal for Academic Development}} (\bibinfo{date}{Aug.} \bibinfo{year}{2022}).
\newblock
\showISSN{1360-144X, 1470-1324}
\urldef\tempurl%
\url{https://doi.org/10.1080/1360144X.2022.2099208}
\showDOI{\tempurl}


\bibitem[Slade and Prinsloo(2013)]%
        {slade2013learning}
\bibfield{author}{\bibinfo{person}{Sharon Slade} {and} \bibinfo{person}{Paul Prinsloo}.} \bibinfo{year}{2013}\natexlab{}.
\newblock \showarticletitle{Learning analytics: Ethical issues and dilemmas}.
\newblock \bibinfo{journal}{\emph{American Behavioral Scientist}} (\bibinfo{year}{2013}).
\newblock


\bibitem[Slade and Tait(2019)]%
        {Slade_Tait_icde-2019}
\bibfield{author}{\bibinfo{person}{Sharon Slade} {and} \bibinfo{person}{Alan Tait}.} \bibinfo{year}{2019}\natexlab{}.
\newblock \bibinfo{booktitle}{\emph{Global guidelines: Ethics in Learning Analytics}}.
\newblock


\bibitem[Steiner et~al\mbox{.}(2016)]%
        {Steiner_Kickmeier-Rust_Albert_2016}
\bibfield{author}{\bibinfo{person}{Christina~M. Steiner}, \bibinfo{person}{Michael~D. Kickmeier-Rust}, {and} \bibinfo{person}{Dietrich Albert}.} \bibinfo{year}{2016}\natexlab{}.
\newblock \showarticletitle{LEA in Private: A Privacy and Data Protection Framework for a Learning Analytics Toolbox}.
\newblock \bibinfo{journal}{\emph{Journal of Learning Analytics}} (\bibinfo{date}{April} \bibinfo{year}{2016}).
\newblock
\showISSN{1929-7750}
\urldef\tempurl%
\url{https://doi.org/10.18608/jla.2016.31.5}
\showDOI{\tempurl}


\bibitem[Tabassi(2023)]%
        {Tabassi-NIST_2023}
\bibfield{author}{\bibinfo{person}{Elham Tabassi}.} \bibinfo{year}{2023}\natexlab{}.
\newblock \bibinfo{booktitle}{\emph{Artificial Intelligence Risk Management Framework (AI RMF 1.0)}}.
\newblock \bibinfo{address}{Gaithersburg, MD}.
\newblock
\urldef\tempurl%
\url{https://doi.org/10.6028/NIST.AI.100-1}
\showDOI{\tempurl}


\bibitem[Tsai and Gasevic(2017)]%
        {Tsai_Gasevic_2017}
\bibfield{author}{\bibinfo{person}{Yi-Shan Tsai} {and} \bibinfo{person}{Dragan Gasevic}.} \bibinfo{year}{2017}\natexlab{}.
\newblock \showarticletitle{Learning analytics in higher education --- challenges and policies: a review of eight learning analytics policies}. In \bibinfo{booktitle}{\emph{Proceedings of the Seventh International Learning Analytics \& Knowledge Conference}}. \bibinfo{publisher}{ACM}, \bibinfo{address}{Vancouver British Columbia Canada}.
\newblock
\showISBNx{978-1-4503-4870-6}


\bibitem[Tsai et~al\mbox{.}(2020a)]%
        {tsai2020privacy}
\bibfield{author}{\bibinfo{person}{Yi-Shan Tsai}, \bibinfo{person}{Alexander Whitelock-Wainwright}, {and} \bibinfo{person}{Dragan Ga{\v{s}}evi{\'c}}.} \bibinfo{year}{2020}\natexlab{a}.
\newblock \showarticletitle{The privacy paradox and its implications for learning analytics}. In \bibinfo{booktitle}{\emph{Proceedings of the tenth international conference on learning analytics \& knowledge}}.
\newblock


\bibitem[Tsai et~al\mbox{.}(2021)]%
        {tsai2021more}
\bibfield{author}{\bibinfo{person}{Yi-Shan Tsai}, \bibinfo{person}{Alexander Whitelock-Wainwright}, {and} \bibinfo{person}{Dragan Ga{\v{s}}evi{\'c}}.} \bibinfo{year}{2021}\natexlab{}.
\newblock \showarticletitle{More than figures on your laptop:(Dis) trustful implementation of learning analytics}.
\newblock \bibinfo{journal}{\emph{Journal of Learning Analytics}} (\bibinfo{year}{2021}).
\newblock


\bibitem[Tsai et~al\mbox{.}(2020b)]%
        {Tsai_Whitelock-Wainwright_=2020}
\bibfield{author}{\bibinfo{person}{Yi-Shan Tsai}, \bibinfo{person}{Alexander Whitelock-Wainwright}, {and} \bibinfo{person}{Dragan Ga\v{s}evi\'{c}}.} \bibinfo{year}{2020}\natexlab{b}.
\newblock \showarticletitle{The privacy paradox and its implications for learning analytics}. In \bibinfo{booktitle}{\emph{Proceedings of the Tenth International Conference on Learning Analytics \& Knowledge}} \emph{(\bibinfo{series}{LAK 20})}.
\newblock
\showISBNx{978-1-4503-7712-6}
\urldef\tempurl%
\url{https://doi.org/10.1145/3375462.3375536}
\showDOI{\tempurl}


\bibitem[Tzimas and Demetriadis(2021)]%
        {Tzimas_Demetriadis_2021}
\bibfield{author}{\bibinfo{person}{Dimitrios Tzimas} {and} \bibinfo{person}{Stavros Demetriadis}.} \bibinfo{year}{2021}\natexlab{}.
\newblock \showarticletitle{Ethical issues in learning analytics: a review of the field}.
\newblock \bibinfo{journal}{\emph{Educational Technology Research and Development}} (\bibinfo{date}{April} \bibinfo{year}{2021}).
\newblock
\showISSN{1042-1629, 1556-6501}
\urldef\tempurl%
\url{https://doi.org/10.1007/s11423-021-09977-4}
\showDOI{\tempurl}


\bibitem[University(2023)]%
        {ou_policy_2023}
\bibfield{author}{\bibinfo{person}{The~Open University}.} \bibinfo{year}{2023}\natexlab{}.
\newblock \bibinfo{title}{Data Ethics Policy 2023}.
\newblock
\newblock
\urldef\tempurl%
\url{https://help.open.ac.uk/documents/policies/ethical-use-of-student-data/files/287/Data%20ethics%20policy%20NEW%20July%202023_.pdf}
\showURL{%
\tempurl}


\bibitem[Veljanova et~al\mbox{.}(2022)]%
        {Veljanova_Barreiros_2022}
\bibfield{author}{\bibinfo{person}{Hristina Veljanova}, \bibinfo{person}{Carla Barreiros}, \bibinfo{person}{Nicole Gosch}, \bibinfo{person}{Elisabeth Staudegger}, \bibinfo{person}{Martin Ebner}, {and} \bibinfo{person}{Stefanie Lindstaedt}.} \bibinfo{year}{2022}\natexlab{}.
\newblock \showarticletitle{Towards Trustworthy Learning Analytics Applications: An Interdisciplinary Approach Using the Example of Learning Diaries} \emph{(\bibinfo{series}{Communications in Computer and Information Science})}.
\newblock
\showISBNx{978-3-031-06390-9}
\urldef\tempurl%
\url{https://doi.org/10.1007/978-3-031-06391-6_19}
\showDOI{\tempurl}


\bibitem[Veljanova et~al\mbox{.}(2023)]%
        {veljanova2023operationalising}
\bibfield{author}{\bibinfo{person}{Hristina Veljanova}, \bibinfo{person}{Carla Barreiros}, \bibinfo{person}{Nicole Gosch}, \bibinfo{person}{Elisabeth Staudegger}, \bibinfo{person}{Martin Ebner}, {and} \bibinfo{person}{Stefanie Lindstaedt}.} \bibinfo{year}{2023}\natexlab{}.
\newblock \showarticletitle{Operationalising Transparency as an Integral Value of Learning Analytics Systems--From Ethical and Data Protection to Technical Design Requirements}. In \bibinfo{booktitle}{\emph{International Conference on Human-Computer Interaction}}. Springer.
\newblock


\bibitem[West et~al\mbox{.}(2016)]%
        {West_Huijser_Heath_2016}
\bibfield{author}{\bibinfo{person}{Deborah West}, \bibinfo{person}{Henk Huijser}, {and} \bibinfo{person}{David Heath}.} \bibinfo{year}{2016}\natexlab{}.
\newblock \showarticletitle{Putting an Ethical Lens on Learning Analytics}.
\newblock \bibinfo{journal}{\emph{Educational Technology Research and Development}} (\bibinfo{date}{Oct.} \bibinfo{year}{2016}).
\newblock
\showISSN{1042-1629}


\end{thebibliography}


\end{document}